%% file: SIGSPATIAL_main.tex
\documentclass[sigconf, authordraft]{acmart}

\usepackage{booktabs} 

\setcopyright{rightsretained}

\acmConference[SIGSPATIAL'17]{ACM SIGSPATIAL}{November 2017}{Los Angeles, California, USA} 
\acmYear{2017}

\begin{document}
\title{Bayesian approach to Spatio-temporally Consistent Simulation of Daily Monsoon Rainfall over India}

\author{Adway Mitra}
\affiliation{
  \institution{International Center for Theoretical Sciences (ICTS-TIFR)}
  \city{Bangalore, India} 
}
\email{adway.cse@gmail.com}

\renewcommand{\shortauthors}{Adway Mitra}
\renewcommand{\shorttitle}{Daily Rainfall Simulation}

\begin{abstract}
Simulation of rainfall over a region for long time-sequences can be very useful for planning and policy-making, especially in India where the economy is heavily reliant on monsoon rainfall. However, such simulations should be able to preserve the known spatial and temporal characteristics of rainfall over India. General Circulation Models (GCMs) are unable to do so, and various rainfall generators designed by hydrologists using stochastic processes like Gaussian Processes are also difficult to apply over the vast and highly diverse landscape of India. In this paper, we explore a series of Bayesian models based on conditional distributions of latent variables that describe weather conditions at specific locations and over the whole country. During parameter estimation from observed data, we use spatio-temporal smoothing using Markov Random Field so that the parameters learnt are spatially and temporally coherent. Also, we use a nonparametric spatial clustering based on Chinese Restaurant Process to identify homogeneous regions, which are utilized by some of the proposed models to improve spatial correlations of the simulated rainfall. The models are able to simulate daily rainfall across India for years, and can also utilize contextual information for conditional simulation. We use two datasets of different spatial resolutions over India, and focus on the period 2000-2015. We propose a large number of metrics to study the spatio-temporal properties of the simulations by the models, and compare them with the observed data to evaluate the strengths and weaknesses of the models.
\end{abstract}




\maketitle

\input{SIGSPATIAL_body}


\end{document}

%% file: SIGSPATIAL_body.tex
\section{Introduction}

Climate conditions are central to the well-being of people, as well as their socio-economic activities. Rainfall is an important aspect of climate, whose importance is enormous in certain parts of the world such as India. Here agriculture is strongly dependent on rainfall as the primary source of irrigation. Various large-scale infrastructures such as dams, reservoirs, roads, bridges etc are also affected by rainfall. So, for planning any agricultural policy or development project, it is important to carry out impact assessment and feasibility studies, using future rainfall as input. Process models, like biophysical crop models and hydrological models for reservoirs require weather data as input. Since actual rainfall data in the future is not available, it is necessary to have simulations of rainfall. However, such simulations must be accurate, and preserve as many of the characteristics of the real rainfall data as possible. But India is a very vast and diverse country, and rainfall over India is quite diverse spatio-temporally. In fact, study of the Indian monsoon is an active research topic in Climate Sciences~\cite{monsoon}.

A large number of climate models of varying levels of complexity have been developed by climate scientists, to simulate meteorological variables worldwide. These are coupled models, which take into account the interactions between many climatic systems and subsystems, using differential equations. A class of such models called General Circulation Models (GCMs) are quite popular, and they provide simulations of rainfall over India, conditioned on simulated climatic conditions all over the world. Some of them have been found to be reasonably accurate in preserving certain properties of Indian Monsoon rainfall, such as inter-annual and intra-seasonal variability~\cite{cmip5}. However, most of the models have been found to be inadequate in capturing other aspects of Indian monsoon, such as its dependence on the Indian Ocean Dipole (IOD). In this work, we find that most of these models are unable to preserve spatio-temporal properties of Indian monsoon rainfall in their simulations.

Due to the inability of GCMs, we now look into another approach:  Stochastic Rainfall Generators. Introduced by C.W. Richardson~\cite{richardson}, they model rainfall occurrence, rainfall volume and sometimes other climatic variables like temperature using conditional probability distributions (as in a Bayesian Network), conditioned on rainfall occurrence. Due to the use of distributions, these methods are capable of simulating the deviation from climatological means. Most of these stochastic simulators follow the general approach of using the training dataset to fit various parameters of these distributions, and then long temporal sequences of meteorological variables are simulated by sampling repeatedly from these distributions. Next, various statistics of interest are computed from this simulated data, and they are compared with the corresponding statistics from the observed data. This is the general approach prescribed by the Intergovernmental Panel on Climate Change (IPCC)~\cite{ipcc} for Stochastic Weather Generators, in which rainfall simulation is the most important step.

Most of the stochastic rainfall generators simulate daily rainfall occurrence (binary) and rainfall volume (real-valued) separately. This is achieved using a latent variable, such as in~\cite{gaussrain}. Temporal coherence is maintained while simulating the rainfall occurrence variable, using Markovian or Semi-Markovian(\cite{poissclus}) approach where the lengths of wet or dry spells are explicitly simulated. Originally location-specific point processes were studied~\cite{pointproc}, but they were unable to capture spatial correlations between neighboring locations, so multi-site processes were introduced~\cite{multisite}. Most of the recently developed stochastic simulators like~\cite{gaussproc,condgen,stlatgauss}achieve spatial correlations by using Gaussian Processes to generate rainfall volume. This approach allows them to make simulations in locations where past data is not available, assuming spatial smoothness over the region. To use Gaussian Process, they need to choose suitable mean and covariance functions. A concise but comprehensive survey on stochastic daily rainfall generators is available in~\cite{stochgensurvey}.

These stochastic rainfall simulations have been used in various parts of the world, such as Argentina~\cite{condgen}, Sweden~\cite{stlatgauss}, USA~\cite{poissclus}, and various countries in Africa~\cite{stochafrica,stochafrica2}. However, not too much work has been done for India, except some attempts like~\cite{northeast}. Simulating Indian monsoon rainfall is highly important and impactful, but also very challenging. Most of the places within the geopolitical boundaries of India receive almost $80\%$ of their total annual rainfall from the South Asian monsoon during June-September. However, some locations such as the south-eastern part of the peninsula remains relatively dry during this period, and receives most of their annual rainfall in the post-monsoon season (October-December). Even among the other locations, the climatological mean rainfall varies quite significantly. There are the dry desert areas in the North-Western part which have only 10-15 days of significant rainfall across the 4 months. On the other hand, the Western coast along Arabian Sea, and some parts of North-eastern India receive heavy rainfall on most of the days. Apart from such spatial diversity, there are also significant intraseasonal daily variations. The first couple of weeks on June are relatively dry, as monsoon onset may not have happened in many of the locations. During the core monsoon months of July, August and September, there are some short sequences of days called ``active spells" when most locations in the country simultaneously receive more rain than their respective means, and other sequences called ``break spells" when most locations in the country are dry. However, during the active spells the North-Eastern areas are usually dry, and during break spells these areas, as well as Northern foothills of the Himalayas and South-eastern parts of the peninsula receive good rainfall. Finally, it is not uncommon that some locations have extremely high rainfall, while some other locations suffer a drought at the same time. Also, the days on which spatial aggregate rainfall across India is maximum, are not necessarily same as the days on which maximum number of locations receive more rainfall than their respective means. A systematic study of all these variabilities is presented in~\cite{monsoon}.

In this work, we aim to build stochastic rainfall simulation models for daily Indian monsoon rainfall, based on latent variables and conditional distributions. However, our approach is quite different from those discussed above. We use binary variables at each spatio-temporal location not to indicate rainfall occurrence or non-occurrence, but to indicate one of two modes corresponding to weather conditions. These modes are associated with heavy rain or light rain, but they may overlap. These variables are expected to be spatio-temporally coherent, as weather type over an area (covering several gridpoints) persists over several days, even though the precipitation amount may vary significantly. The  Distributions on these variables are learnt from training data, ensuring their spatio-temporal coherence through a Markov Random Field. We also use a variable which indicates the all-India condition (such as active spells, break spells, and normal spells). We do not use Gaussian process to simulate rainfall volumes at locations, because the covariance function is difficult to construct over the heterogeneous landmass. Most covariance functions used for Gaussian Processes imply that correlations between rainfall volumes at two locations are strongly correlated to their geographical (Euclidean) distance. However, this is not the case in India, where locations on the western and eastern slopes of the Western Ghats mountin range (that runs along the western coast) are less than 100 Km away, but their rainfall characteristics are completely different as the mountain range creates a rain shadow zone. Instead, we attempt to demarcate the landmass into a suitable number of homogeneous zones where all locations are in the same mode (as mentioned earlier) on most of the days of the training period. Some of our models take account of these zones to ensure spatial coherence.

Finally, we discuss evaluation criteria for the simulation models while comparing the statistics from the simulated data to those of the observed data. We lay special emphasis on how well spatio-temporal properties of the observed data are preserved in the simulation. We study the mean correlation of the daily rainfall at each location with those at its neighboring locations, and also the  spatial distribution of rainfall across the locations. We study the mean correlation of rainfall amounts at all locations on pairs of successive days, and also the mean lengths of dry and wet spells at all locations.  Finally, we also evaluate the mean and standard deviation of daily rainfall for each location, and also for the all-India spatial aggregate. We evaluate our proposed models by generating simulation data from them conditioned on some input, like rainfall observations at a few random locations on random days, or the all-India aggregate rainfall per day.We also evaluate many General Circulation Models (listed in~\cite{cmip5}) with these criteria, and point out their glaring weaknesses.

The main contributions of this work, relative to the existing literature on stochastic daily rainfall generators are as follows: 1) We build simulators for a vast and diverse spatial field like India, unlike most simulators that are built for relatively small and homogeneous regions, 2) We make use of Markov Random Fields to ensure spatio-temporal coherence while learning location-specific distribution parameters for the latent variables, 3) We model country-wide broad weather type (like active/break spells) and their relations with individual locations, 4) We use a nonparametric clustering approach to identify homogeneous sets of locations where local weather conditions are same on most days, 5) We study different ways of incorporating supervisions to improve our simulations, 6) We come up with many novel measures to quantify and compare spatio-temporal properties of the simulations, and finally 7) Instead of a single model, we make a series of models and show their relative merits and demerits with respect to these measures. We even evaluate simulations by GCMs using these measures.

\section{Datasets, Variables and Parameters}
For this work, we use two datasets, both compiled and released by Indian Meteorological Department (IMD). In the first dataset, daily rainfall data from 1901 to 2011 is available over 357 grid-points all over India, each grid-point of size 100Km-by-100Km. In the second dataset, daily rainfall data is available for the months April-November, from 1901 to 2014, over 4964 grid-points all over India, each of which is of size 25Km-by-25Km. Since most of the places in India receive almost $80\%$ of their annual rainfall from the Indian monsoon (June-September), in this work we use only these four months for our simulation. However, during this period the south-eastern parts of the peninsula remains relatively dry. Also, we focus on the period 2000-2014 instead of the entire duration since 1900, because climate change has resulted in various changes in Indian monsoon characteristics across this period, and so the parameters cannot be considered as constant.

For modelling purposes, we now introduce the notations and variables. Suppose there are $S$ locations, and the total number of days is $T$. Any location $s$ has a set of neighboring locations $NB(s)$, according to the grid coordinates. Only locations lying on Indian geo-political landmass are considered. At each location $s$ and day $t$, $X(s,t)$ denotes the volume of rainfall received, while $Y(t)$ denotes the aggregate rainfall received by the entire country on that day. When these variables are measured from the dataset, we denote them as $X^{DATA}(s,t)$ and $Y^{DATA}(t)$. When we consider simulation outputs by a model $M$, they are denoted as $X^{M}(s,t)$ and $Y^{M}(t)$.

Now, we introduce two latent variables that indicate the rainfall conditions. Each state of binary variable $Z(s,t)$ represents a distribution over the rainfall volume at location $s$ and day $t$, one state $(Z=1)$ peaked at higher value and the other $(Z=2)$ close to 0. In other words,
\begin{equation}
X(s,t) \sim Gamma(\alpha_{skt},\beta_{skt}) \textit{ where } k=Z(s,t)
\end{equation}
where $(\alpha,\beta)$ are the parameters of a Gamma distribution dependent on $Z$, and potentially varying across locations and time. However, in this work we drop the time-dependence of these parameters to improve model complexity. This is somewhat similar to the rainfall occurrence variable considered by most stochastic rainfall generators such as \cite{condgen,gaussrain}, but not exactly same. $Z_{st}$ actually corresponds to the weather condition at location $s$ and day $t$, that is expected to be spatio-temporally coherent. Ideally, $Z$ should be based not only on rainfall but also other meteorological variables such as cloud cover that influence rainfall. At each location $s$, we use a distribution $prob(Z(s,t)=k|Z(s,t-1)=l)=\tau_{slk}$ that quantifies the temporal coherence of $Z$.

We also consider a variable $U(t)$ that takes 3 values and indicates the rainfall conditions over the entire country. $U=1$ is associated with active spells~\cite{monsoon}, and signifies that most of the $S$ locations are in state $Z=1$. But $U=2$ is associated with the pre-onset and break spells~\cite{monsoon}, and signifies that most of the $S$ locations are in state $Z=2$. $U=3$ signifies normal conditions. $U$ is also expected to be temporally coherent. For each location $s$, its relation to the all-India condition is encoded by a distribution $prob(Z(s,t)=k|U(t)=l)=\theta_{slk}$. For most locations, $U=1$ implies higher chance of $Z=1$ than $Z=2$, but for some locations mostly in the north-eastern parts of the country and parts of the eastern coast, it is the reverse~\cite{monsoon}.

\section{Parameter learning by Markov Random Fields}
Clearly, for simulation we need to learn the parameters $\alpha,\beta,\theta$ and $\tau$, for which we must know the latent variables $Z$ and $U$. In the model training phase, we infer these state variables. The naive way to do so is to consider the rainfall time-series at each location independently, and use Expectation-Maximization approach to fit a 2-state Hidden Markov Model with Gamma emission distribution, and assign $Z$-variables accordingly. However, this does not account for spatial coherence of the $Z$-variables. Also, it is less easy to do this for $U$ variables, since each value of $U(t)$ has a bearing on the number of locations that are in state $Z=1$ on day $t$, and it is difficult to fit a distribution on this number. So, we make use of a Markov Random Field to find the best assignment of $Z$ and $U$ variables that fits the observations $X$ and also preserves spatio-temporal coherence.

In the Markov Random Field, we have two nodes $(Z(s,t),X(s,t))$ for each spatio-temporal location $(s,t)$ where $s \in \{1,S\}$ and $t \in \{1,T\}$, and $T$ is the length of the training sequences. We also have the all-India state node $U(t)$ for each day. Each $Z(s,t)$ node is connected by \emph{temporal edges} to $Z(s,t+1)$ and $Z(s,t-1)$, and by \emph{spatial edges} to $Z(s',t)$ where $s'$ is a spatial neighbor of $s$, i.e. $s' \in NB(s)$. $Z(s,t)$ is also connected to $U(t)$ by \emph{scale edge} and to $X(s,t)$ by \emph{data edge}. Each $U(t)$ is also connected to $Y(t)$ by data edges, and to $U(t+1),U(t-1)$ by temporal edges. The graphical model is shown in Figure 1.
\begin{figure}
	\centering
	\includegraphics[width=3.5in,height=2in]{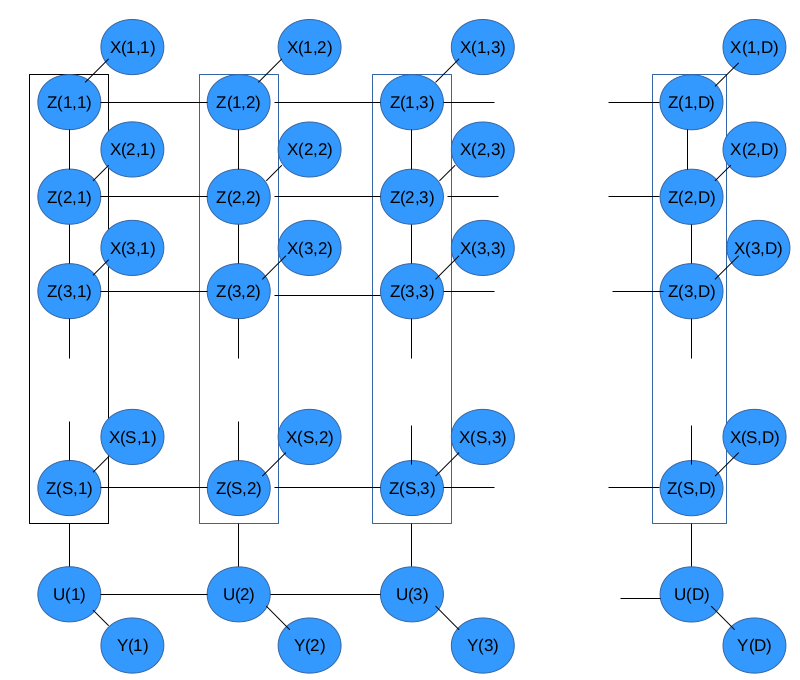}
	\caption{Markov Random Field as mentioned above. The horizontal edges are temporal edges, vertical edges are spatial edges, diagonal edges are data edges. All the $Z$-variables each day are linked to the $U$ variable of that day.}
\end{figure}
On each of these edges, we define \emph{potential functions} $\Psi^e$. For any spatial, temporal or scale edge $e$, we define the function such that it takes a high value $a^e$ if the nodes connected by it take same value, and low value $b^e$ if they take different values. For spatial edges between any pair of neighboring locations $(s,s')$, we set $a^e$ equal to the correlation between $X^{DATA}(s)$ and $X^{DATA}(s')$ across the temporal duration $T$ of the training sequence, and $b^e$ is set to 0. For any scale edge from location $s$, $a^e$ is set to exponential of the correlation between $X(s)$ and $Y$ across the $T$ time-points, and $b^e$ is 1. For all temporal edges on $Z$-variables of a location $s$, $(a^e,b^e)$ are constants, and we set to their ratio to 99. This corresponds to the prior distribution of $prob(Z(s,t)=Z(s,t-1))=0.99$. For data edges between $Z(s,t)$ and $X(s,t)$, we define the potential function according to Equation 1, i.e. PDF at $X(s,t)$ of the Gamma distribution whose parameters are specified by $Z(s,t)$. For data edges between $U(t)$ and $Y(t)$, we set the potential function equal to the PDF at $Y(t)$ of a Gaussian distribution whose parameters $(\mu,\sigma)$ are specified by $U(t)$. The likelihood of the assignment of latent variables $(Z,U)$ and the parameters $(\alpha,\beta,\mu,\sigma)$ conditioned on $X$ and $Y$ is the product of all these potential functions. Mathematically, for $Z$ the distribution is

\begin{small}
	\begin{eqnarray}
	\mathcal{L}(Z,U,\alpha,\beta,\mu,\sigma|X^{DATA}) \propto \prod_{s,t}\prod_{s'\in NB(s)}\Psi^{spatial}(Z(s,t),Z(s',t)) \nonumber \\ 
	\times \prod_{s,t}\Psi^{temporal}(Z(s,t),Z(s,t-1)) \times \prod_{t}\Psi^{temporal}(U(t),U(t-1))	 \nonumber \\
	\times \prod_{t}\Psi^{data}(U(t),Y^{DATA}(t)) \times \prod_{s,t}\Psi^{data}(Z(s,t),X^{DATA}(s,t))  \nonumber\\ 
	\times \prod_{s,t}\Psi^{scale}(Z(s,t),U(t))) 
	\end{eqnarray}
\end{small}

Note that the use of exponential on the potential functions of scale edges ensures that the all-India weather state $U(t)$ depends on local weather states $Z(s,t)$ at all the locations on each day, without defining any conditional distribution. It also depends on the aggregate all-India rainfall $Y(t)$. 

The next task is to estimate these unknown quantities such that this likelihood function is maximized. Clearly, the spatial, temporal and scale potential functions encourage the $Z$ and $U$ variables to match their spatio-temporal neighbors, though they also need to fit the data, thereby requiring a compromise. For an initial estimate, we neglect all spatial, temporal and scale edges, and independently estimate the unknown variables and parameters using Expectation Maximization. Clearly, such an estimate is not spatio-temporally coherent. To achieve that end, we now consider the full graph structure again, and carry out inference using Gibbs Sampling, where in each step we sample one $Z(s,t)$ or one $U(t)$ variable conditioned on all the remaining $Z$ and $U$ values. The parameters $(\alpha,\beta,\mu,\sigma)$ are also updated accordingly. Gibbs Sampling is relatively straightforward, since each $Z(s,t)$ is conditionally independent of all $Z$ and $U$ given its neighoring vertices (a key property of MRF).

\begin{small}	
	\begin{eqnarray}
	&pr(Z(s,t)=k|Z,U,X^{DATA},Y^{DATA})  \nonumber \\
	&= pr(Z(s,t)=k|Z(s,t+\delta),Z(s',t),U(t),X^{DATA}(s,t)) \nonumber\\
	&\propto pr(Z(s,t)=k,Z(s,t+\delta),Z(s',t),U(t),X^{DATA}(s,t))  \nonumber\\
	&=\prod_{\delta\in(-1,1)}\Psi^{temporal}(Z(s,t)=k,Z(s,t+\delta))   \nonumber\\
	&\times\prod_{s'\in NB(s)}\Psi^{spatial}(Z(s,t)=k,Z(s',t)) \nonumber\\
	&\times\Psi^{scale}(Z(s,t)=k,U(t))\times\Psi^{data}(Z(s,t)=k,X^{DATA}(s,t))
	\end{eqnarray}
\end{small}

\begin{small}
	\begin{eqnarray}
	&pr(U(t)=k|Z,U,X^{DATA},X^{DATA}) \propto pr(U(t)=k,U(t+\delta),Z(t),Y^{DATA}(t)) \nonumber\\
	&= \prod_{\delta\in(-1,1)}\Psi^{temporal}(U(t)=k,U(t+\delta)) \nonumber\\
	&\times\prod_s^S\Psi^{scale}(Z(s,t),U(t)=k)\times\Psi^{data}(U(t)=k,Y^{DATA}(t))	
	\end{eqnarray}		
\end{small}

After performing these samplings iteratively and collecting samples at regular intervals, we find the mode of $(Z,U)$, along with updated estimates of the parameters $(\alpha,\beta,\mu,\sigma)$. This estimate of $Z^{MRF}$ and $U^{MRF}$ also allow us to make MAP estimate of the parameters $\tau$ and $\lambda$ (state transition distributions of $Z$ at each location, and $U$) and $\theta$ (distribution of $Z$ at each location, conditioned on $U$). As $Z^{MRF}$ and $U^{MRF}$ estimated by the MRF are spatio-temporally coherent, the posterior estimate of the parameters too reflect this property. We also estimate a posterior distribution on $U$.

\section{Simple Simulation Models}
The training process is completed using the MRF-based MAP estimation of the parameters $(\alpha,\beta,\tau,\theta)$, and an estimation of the latent variables $(Z,U)$ which we can use as ``ground truth" for evaluating our simulations. Now, we are ready to build the models for simulation. In this section, we will introduce four simple models, two of which are single-site models, while the other two make use of interaction between local and all-India weather states. Although these models are simplistic compared to the state-of-the-art models, they are important for the comparison of properties that will follow.

\subsection{Model 1}
The first model is a simple single-site model, along the lines of~\cite{pointproc}. Here, each $Z^{M1}(s,t)$ is sampled independently, followed by $X^{M1}(s,t)$. Mathematically,
\begin{small}
\begin{eqnarray}
Z^{M1}(s,t)\sim Bernoulli(\hat{\tau}_s); X^{M1}(s,t)\sim Gamma(\alpha_{sk},\beta_{sk}) \nonumber  \\
\textbf{ where } k=Z^{M1}(s,t); \forall s\in\{1,S\},t\in\{1,T\} \nonumber
\end{eqnarray}
\end{small}
Here, $\hat{\tau}$ is the marginal distribution of the states computed from the state-transition distribution $\tau$.

This model requires 5 parameters $\alpha_{s1}$,$\beta_{s1}$,$\alpha_{s2}$,$\beta_{s2}$,$\hat{\tau}_s$ for each location, i.e. totally $5S$ parameters.

\subsection{Model 2}
The second model is also single-site, but this time we consider the temporal dynamics of $Z^{M2}(s,t)$ as a Markov process. Mathematically,
\begin{small}
\begin{eqnarray}
&Z^{M2}(s,1)\sim\hat{\tau}_s; X^{M2}(s,1)\sim Gamma(\alpha_{sk},\beta_{sk}) \nonumber \\
&Z^{M2}(s,t)\sim{\tau}_{sl}; X^{M2}(s,t)\sim Gamma(\alpha_{sk},\beta_{sk}) \nonumber \\
&\textbf{ where } l=Z^{M2}(s,t-1); k=Z^{M2}(s,t); \forall s\in\{1,S\},t\in\{2,T\} \nonumber
\end{eqnarray}
\end{small}

This model requires 6 parameters $\alpha_{s1}$, $\beta_{s1}$, $\alpha_{s2}$, $\beta_{s2}$, $\tau_{s1}$, $\tau_{s2}$ for each location, i.e. totally $6S$ parameters.

Clearly, the use of Markov process will make the simulations from this model temporally coherent, and dry/wet spells can be simulated with their lengths following geometric distribution. Some stochastic generators such as~\cite{poissclus} have done away with the Markov approach of achieving this and used a semi-Markov approach where the lengths of these spells are modelled explicitly as Poisson distribution with parameters specific to locations and states. But we could not use this approach since the distribution of lengths of such spells are so different in different locations over India that no single type of distribution can be used. Using different types of distributions in different locations will make the model too ugly, which we wanted to avoid.

\subsection{Model 3}
In the third model, we consider the relations between all-India weather state variable $U$ and the local weather state variable $Z$. For each day, first the $U$ variables are sampled. This can be done according to a state transition distribution $\lambda$ for $U$, learnt from $U^{MRF}$. But they can also be estimated from supervision information, which allows us to perform conditional simulation based on  coarse external information that acts as a driver (like~\cite{condgen}).
\begin{small}
\begin{eqnarray}
&U^{M3}(1) \sim \hat{\lambda}; U^{M3}(t) \sim \lambda_n \textbf{ where } n=U^{M3}(t-1) \nonumber \\
&Z^{M3}(s,t) \sim \theta_{sl}; X^{M3}(s,t)\sim Gamma(\alpha_{sk},\beta_{sk}) \nonumber \\
&\textbf{ where } l=U^{M3}(t); k=Z^{M3}(s,t); \forall s\in\{1,S\},t\in\{1,T\} \nonumber
\end{eqnarray}
\end{small}

This model requires 7 parameters $\alpha_{s1}$,$\beta_{s1}$,$\alpha_{s2}$,$\beta_{s2}$,$\theta_{s1}$,$\theta_{s2}$,$\theta_{s3}$ for each location, and additionally 6 parameters for $\lambda$ i.e. totally $7S+6$ parameters.

Since the MRF-based parameter estimation has ensured that the parameters are spatially correlated, we can expect to see an increased spatial coherence of $Z^{M3}$ and spatial correlation of $X^{M3}$ under the driving effect of $U$.

\subsection{Model 4}
Finally, in model 4 we combine models 2 and 3 together, by defining conditional distributions $\pi_s$ that denote $prob(Z^{M4}(s,t)=k|Z^{M4}(s,t-1)=l,U^{M4}(t)=m) = \pi_{slmk}$. These are also learnt a-posteriori using $Z^{MRF}$ and $U^{MRF}$. The model is as follows:
\begin{small}
\begin{eqnarray}
&U^{M4}(1) \sim \hat{\lambda}; U^{M4}(t) \sim \lambda_n \textbf{ where } n=U^{M4}(t-1) \nonumber \\
&Z^{M4}(s,t) \sim \pi_{slm}; X^{M4}(s,t)\sim Gamma(\alpha_{sk},\beta_{sk}) \nonumber \\
&\textbf{ where } m=U^{M4}(t); l=Z^{M4}(s,t-1); k=Z^{M4}(s,t); \nonumber \\
&\forall s\in\{1,S\},t\in\{1,T\} \nonumber
\end{eqnarray}
\end{small}

This model requires 10 parameters $\alpha_{s1}$, $\beta_{s1}$, $\alpha_{s2}$, $\beta_{s2}$, $\pi_{s11}$, $\pi_{s12}$, $\pi_{s13}$, $\pi_{s21}$, $\pi_{s22}$, $\pi_{s23}$ for each location, and additionally 6 parameters for $\lambda$ i.e. $10S+6$ parameters totally.

This model hopes to achieve temporal coherence by conditioning on $Z(s,t-1)$ and spatial coherence by conditioning on $U(t)$.

\section{Identification of Coherent Zones}
To reduce the parameter complexity, as well as improve spatial coherence, we now attempt to partition the landmass into coherent zones,  so that some of the model variables such as $Z$ can be made specific to zones rather than to locations. In the literature, various attempts at regionalization of the Indian landmass has been made based on rainfall characteristics~\cite{indclus}, but these are mostly with respect to annual statistics. In this work, we are more interested in identifying sets of locations where each of them can be assigned the same value of $Z$ every day. For this purpose we use the $Z^{MRF}$  assignments into the framework of spatial clustering. The $T$-dimensional binary vector $Z^{MRF}_s$ from each location $s$  serves as the set of feature vectors. However, since we do not know the number of clusters, i.e. coherent zones to be formed, we cannot use approaches like Spectral Clustering. Instead, we make use of Nonparametric approaches based on Chinese Restaurant Process. Such methods have been used for spatial clustering, in context of image segmentation ~\cite{spatCRP, spatCRP2}.

Consider each locations $s$ is assigned to a coherent zone $H(s)$. Also consider a set $V$ of canonical binary vectors $\{V_1,V_2,\dots\}$ of dimension $T$, each of which corresponds to the $Z$-vectors for a coherent zone. The $Z$-vector of each location is a somewhat corrupted version of $V_{H(s)}$, where an expected fraction $p$ of the binary entries are flipped, i.e. on an expected number $Tp$ of all the $T$ days, the local weather state at any location is different from the weather state of its corresponding zone. The number of zones to be created clearly depends on $p$, let this number be $K_p$. 

Now, we introduce the generative model based on Spatially Coherent Chinese Restaurant Process (SC-CRP) on this setting. For each location $s$, we assign to it a zone id $H(s)$, which can be among the zones assigned to the neighboring locations, or a separate zone. This ensures that all the zones are spatially coherent; no location is assigned to a zone unless at least one of its neighboring  locations is also assigned to that zone, or it is a single-point zone. As with normal Chinese Restaurant Process, if we consider the assignment process sequentially, the probability of assigning any location $s$ to a zone $k$ is proportional to the number of locations $n_k$ already assigned to it, and that of assigning $s$ to a new coherent zone is proportional to a constant $\alpha$. Once this has been done, the binary $Z$-vector for that location $s$ is generated by flipping each of the elements of $V_{H(s)}$ with a probability $p$. The generative model based on SC-CRP can be written as follows:

\begin{small}
\begin{eqnarray}
&prob(H(s) = k | H(1,\dots,s-1)) \propto n_k \textnormal { if } \exists s'\in NB(s) \textnormal{ s.t. } H(s')=k \nonumber \\
&\propto \alpha \textnormal{ if } \not \exists s' \textnormal{s.t.} H(s')=k  \nonumber \\
&= 0  \textnormal{ otherwise } \nonumber \\
&Z(s,t) \sim Ber(V(z,t),p) \textnormal{ where } z=H(s), t \in \{1,T\}, s \in \{1,S\}  \nonumber
\end{eqnarray}
\end{small}

However we know only realizations of $Z$, in the form of $Z^{MRF}$, and not $H$ and $V$. So, we use Gibbs Sampling to perform the inference on $H$, with $V$ re-estimated with each iteration. Finally we get $K_p$ coherent zones, which depends on $p$. Naturally, if $p=1$ then $K_p=S$. We find that for the low-resolution dataset with $S=357$, the number of zones is 129 for $p=0.9$, while for the high-resolution dataset with $S=4964$, we have $K_{0.9}=248$. 

\section{Models for Spatially coherent Simulation}
In this section we propose our remaining two models, which are along the lines of the previous models but use the spatially coherent zones identified above. This allows us to gain additional spatial coherence of $Z$.

\subsection{Model 5}

This model is along the lines of Model 4, but using an additional variable $C$ for weather state at zone $z$. The $\pi$ distributions are now defined over these zones instead of locations. Once the zonal weather states $C$ have been simulated according to $\pi$, the local weather states $Z(s,t)$ are selected by setting them equal to the corresponding zonal state $C(H(s),t)$ with probability $p$, and the reverse of the zonal state with probability $(1-p)$. This is done according to the Chinese Restuarant Process model of spatial clustering discussed above. The model is as follows:

\begin{small}
\begin{eqnarray}
&U^{M5}(1) \sim \hat{\lambda}; U^{M5}(t) \sim \lambda_n \textbf{ where } n=U^{M5}(t-1) \nonumber \\
&C^{M5}(z,t) \sim \pi_{zlm}; \forall z\in\{1,K_p\},t\in\{1,T\} \nonumber \\
&\textbf{ where } m=U^{M5}(t); l=C^{M5}(z,t-1); \nonumber \\
&Z^{M5}(s,t) \sim Ber(c,p); X^{M5}(s,t)\sim Gamma(\alpha_{sk},\beta_{sk}) \nonumber \\
&\textbf{ where } c=C^{M5}(H(s),t); k=Z^{M5}(s,t); \forall s\in\{1,S\},t\in\{1,T\} \nonumber
\end{eqnarray}
\end{small}

In this model, each location $s$ requires 5 parameters $\alpha_{s1}$, $\beta_{s1}$, $\alpha_{s2}$, $\beta_{s2}$, $H(s)$ while each zone $z$ requires 6 parameters $\pi_{z11}$, $\pi_{z12}$, $\pi_{z13}$, $\pi_{z21}$, $\pi_{z22}$, $\pi_{z23}$, 6 parameters for $\lambda$ and finally there is $p$. So totally there are $5S+6K_p+7$ parameters.

\subsection{Model 6}

Finally, we come to our final model, which is an extension of Model 5, but with the express purpose of scaling up spatial correlation of local rainfall volume $X(s,t)$ with its neighbors. Unlike all the previous models, here the rainfall amount sampled as Gamma distribution is not specific to locations but to the coherent zones, conditioned on the zonal weather state $C$ as in Model 5, and denoted by $W(z,t)$. This rainfall volume is distributed among the constituent locations of each zone, according to a distribution $\phi(z)$, learnt from the training dataset. 

\begin{small}
\begin{eqnarray}
&U^{M6}(1) \sim \hat{\lambda}; U^{M6}(t) \sim \lambda_n \textbf{ where } n=U^{M6}(t-1) \nonumber \\
&C^{M6}(z,t) \sim \pi_{zlm}; W^{M6}(z,t)\sim Gamma(\alpha^C_{zk},\beta^C_{zk}) \nonumber \\
&\textbf{ where } m=U^{M6}(t); l=C^{M6}(z,t-1); k=C^{M6}(z,t); \nonumber \\
&\forall z\in\{1,K_p\},t\in\{1,T\} \nonumber \\
&X^{M6}(s,t)=\phi(H(s),s)W^{M6}(H(s),t) \forall s\in\{1,S\},t\in\{1,T\} \nonumber
\end{eqnarray}
\end{small}
Note that $Z(s,t)$ is not separately assigned in this model. For evaluation purposes in the next section, we will consider $Z^{M6}(s,t)=C^{M6}(H(s),t)$.

In this model, each zone requires 10 parameters $\alpha_{z1}$, $\beta_{z1}$, $\alpha_{z2}$, $\beta_{z2}$, $\pi_{z11}$, $\pi_{z12}$, $\pi_{z13}$, $\pi_{z21}$, $\pi_{z22}$, $\pi_{z23}$, while each location needs two parameters $(H(s),\phi(H(s),s))$, 6 parameters for $\lambda$ and additionally there is $p$. Totally there are $10K_p+2S+7$ parameters. Although $K_p$ depends on the choice of $p$, for all practical purposes this is the most concise model.

Use of the distribution $\phi(z)$ implies that proportion of rainfall received by the locations within a zone is the same on all days. This is unrealistic. A small trick to prevent it is to sample an uniform Dirichlet-distributed PMF $\gamma(z,t) \sim Dir(r)$ for each day, and use it to corrupt $\phi(z)$ as $\hat{\phi}(z,t)=q\phi(z)+(1-q)\gamma(z,t)$, where $q$ is a suitably chosen value, probably in the range $(0.7-0.9)$. This $\hat{\phi}(z,t)$ can be used in Model 6 now.

\section{Conditional Simulation}
Now that the models are ready, we consider various settings for the simulations to run. One possibility is for the simulations to run unconditionally, i.e. without any external input apart from the parameters. In case of such a simulation, all the latent variables are simulated according to the models as described above.  Such a simulation can give us an estimate for the future as it runs fully independently, and hence it is useful for impact assessment etc. However, since it has no link with the actual conditions of a specific year, a comparison with the data (once it becomes available) or across different models for their evaluation is almost meaningless. The only way to evaluate unconditional simulations is to let them run long enough  and compare the long-term statistical properties of such simulations with those of the data (when it become available).

The other alternative is \emph{conditional simulation}, like~\cite{condgen}, which runs conditioned on some information about the days or years being simulated. Usually, this information is quite coarse level, but we may use them to estimate some of the random variables, and this estimation has an impact on the simulation of the other variables. We consider two types of conditions in this work: the daily all-India rainfall; and local rainfall at a random set of locations and days.

In the first case, the total rainfall over India, i.e. $Y(t)$ is known for every day. We use this information to infer $U(t)$, and this is in turn used as input to the models. Models 1 and 2 cannot make use of it, but Models 3-6 are benefitted from it. The inference of $U$ is done by considering a Hidden Markov Model with 3 states having Gaussian emission with parameters $\{\mu_k,\sigma_k\}_{k=1}^3$ and state transition distribution $\lambda$ - all estimated from $U^{MRF}$ and $Y^{DATA}$. 

In the second case, the local rainfall $X$ in each of the total $ST$ spatio-temporal locations, is made known with a probability $p$, i.e. $X$ is known in about $pST$ spatio-temporal locations. Based on these, we make an estimate of all $Z$-variables and then the $U$-variables. In each of the locations say $s$, an estimate of the $Z$-variables in each of the ``observed" days (i.e. where $X^{DATA}(s,t)$ is known) is made, using $(\alpha,\beta)$ parameters. After this, the $Z$-variables for location $s$ are estimated in the remaining days as well, using $\tau$ by an iterative process. In each step of the iteration, $Z$ is estimated for any day $t$ if at least one of $Z(s,t-1)$ and $Z(s,t+1)$ has already been estimated in the previous iterations. This process continues till $Z$-estimates have propagated to all days, for those locations that have at least one observation. Once an estimate of $Z$ is made in all the locations and days, an estimate of $U$ for all the days are made, using $\theta$. These $U$-estimates are then used to drive the models once again, as in the previous case. Note that Model 1 cannot benefit from this supervision (as all the spatio-temporal locations are independent for it), while Model 2 makes use of the estimates of $Z$. Obviously, the simulation will be closer to the true data for higher values of $p$.

\section{Performance Measures}
Now, we discuss the measures by which we compare the simulation results to the data. This step is most essential to understand the strengths and weaknesses of each model. We compare both location-specific properties, and spatio-temporal patterns. There are two categories of evaluation measures: for latent variables ($Z$,$U$) and for observed variables ($X$,$Y$).

\subsection{Properties of Latent Variables}
When the MRF is run for a particular duration, say a year, and the learnt parameters are used to simulate the same year, we should compare the simulated latent variables for model verification. For this purpose, we set certain criteria based on these variables to evaluate the simulations. First of all we have \emph{state bias} $ZZ1$. i.e. the number of spatio-temporal locations (out of $ST$) having $Z=1$. This helps to understand if the model is biased towards any state. Next we have \emph{spatial coherence} $Scoh$: on each day, the mean fraction of neighbors of any location $s$ that have the same $Z$-value as $s$. The mean is also taken across all days. Similarly, we evaluate \emph{temporal coherence} $Tcoh$: for any location, the mean fraction of days that it has same $Z$-value as the previous day. The mean is also taken across all locations. Note that this is related to $\tau$ and $\pi$ parameters.

The Indian landmass has a lot of spatial diversity of daily rainfall - even if the daily spatial aggregate rainfall is very high on a given day, not all locations may have significant rainfall on that day. Some locations receive very high rainfall during the break phases, while some locations remain dry during the active phases~\cite{monsoon}. We measure this \emph{spatial diversity} using three indices: $SpDiv$ - the correlation between daily all-India rainfall $Y$ and the number of locations in state $Z=1$ in each day; $nZ1U1$ - the mean number of locations in state 1 on days when $U=1$, and $nZ1U2$ - the mean number of locations in state 1 on days when $U=2$. Note that these are related to $\theta$ and $\pi$ parameters.

\subsection{Properties of Observed Variables}
The latent variables are important for the models, but from an application point of view, the observed variables $X$ and $Y$ are of prime importance. So we now set criteria to evaluate simulations based on these variables. First of all, we standardize the values of $X$ and $Y$ in all models with respect to the mean of all local $X$-variables. This is not necessary for our proposed models, but important for the GCMs since some of them produce transformed values. 

We compute the mean and standard deviation of $X$ at all locations, and also $Y$. We compare $SY$- the standard deviation of daily all-India rainfall (mean all-India rainfall is similar for most models after the standardization). Since we cannot tabulate the location-specific statistics of all locations, we instead compute $dMX$ and $dSX$: the mean relative error in these quantities, i.e. $dMX=mean_{s}\frac{|mn^{MODEL}_s(X)-mn^{DATA}_s(X)|}{mn^{DATA}_s(X)|}$ where $mn_s(X)$ is the mean of $X$ at location $s$ across all the days, and $dSX=mean_{s}\frac{|sd^{MODEL}_s(X)-sd^{DATA}_s(X)|}{sd^{DATA}_s(X)|}$  where $sd_s(X)$ is the standard deviation of $X$ at location $s$ across all the days. Also, to see how well \emph{local extreme rainfall} are simulated, we also measure $X100$: the total number of times that any location has received over 100mm of rainfall on any day. 

Next, we come to mean lengths of wet spells $wetln$, the mean number of successive days that a location receives over 10mm of rainfall. Once again, the mean is taken across all locations. Then we have three measures based on correlations. First, we have \emph{daily correlation} $dcr$ - the correlation between $Y^{DATA}$ and $Y^{MODEL}$ across all the days in the simulated period. We evaluate $scr$ -\emph{mean spatial correlation} -of each location with its neighboring locations on same day. The mean is computed across locations and days. Next we have \emph{correlations of spatial patterns} - $S$-dimensional vector of rainfall volume at each location. We compute this pattern each day and compute its correlation with the pattern of the previous day, and the mean correlation across all days is evaluated as $tcr$. Again, we compute the mean spatial pattern across all the days for both the data and the simulation, and evaluate their correlation as $spatcr$.

\section{Evaluation of GCMs}
Before evaluating the proposed models, we first present our evaluations for various General Circulation Models. In particular, we focus on the ones identified by~\cite{cmip5} as reasonably successful in simulating certain aspects of Indian monsoon rainfall. Since for most models we have data only till 2005, we focus on the period 2000-2005. Also, most of these models operate at coarser resolution than $100KM-100KM$, which is the resolution of our low-resolution dataset with $S=357$ locations. So we downscaled the GCM observations to the same grid with $S=357$ locations. The evaluation criteria for observed variables as discussed above are used for this evaluation. However, we do not evaluate spatial correlation because of the spatial smoothening done. 

\begin{table}
\centering
\begin{tabular}{| c | c | c | c | c | c | c | c | c |}
\hline
Model & dMX & dSX & SY & X100 & wetln & dcr & tcr & spatcr \\
\hline
DATA  & 0   & 0   & 1230 & 1300 & 1.9 & 1 & 0.37 & 1\\
MIROC5 & 0.34 & 0.38 & 923 & 598 & 2.7 & 0.3 & 0.64 & 0.71\\
CCSM4  & 0.48 & 0.21 & 1054 & 498 & 2.6 & 0.2 & 0.53 & 0.69\\
BCC    & 0.23 & 0.41 & 1710 & 2459 & 2.1 & 0.15 & 0.41 & 0.5\\
BNU-ESM& 0.32 & 0.48 & 1045 & 9 & 3.8 & 0.28 & 0.8 & 0.5\\
CESM-BGC& 0.35 & 0.3 & 1026 & 696 & 2.4 & 0.23 & 0.58 & 0.68\\
CESM-CAM5& 0.22 & 0.46 & 853 & 539 & 3.9 & 0.35 & 0.62 & 0.65\\
CMCCCM5 & 0.65 & 0.55 & 1524 & 854 & 1.8 & 0.11 & 0.62 & 0.66\\
CNRMCM5 & 0.27 & 0.49 & 1314 & 1376 & 3.4 & 0.16 & 0.64 & 0.69\\
GFDLCM3 & 0.54 & 0.4 & 997 & 144 & 5 & 0.1 & 0.72 & 0.44\\
GFDLESM2G& 0.65 & 0.67 & 980 & 543 & 2.8 & 0.17 & 0.78 & 0.56\\
HADCM3 & 0.53 & 0.59 & 1768 & 171 & 4 & 0 & 0.81 & 0.6\\
HADGEM2 & 0.78 & 0.69 & 1385 & 1831 & 2.3 & 0 & 0.75 & 0.48\\
IPSL-CM5 & 0.38 & 0.52 & 1573 & 1203 & 3.1 & 0.04 & 0.72 & 0.33\\
MIROC-ESM & 0.55 & 0.32 & 1401 & 35 & 6.6 & 0.02 & 0.65 & 0.51\\
MPI-ESM & 0.52 & 0.42 & 1391 & 802 & 2 & 0.25 & 0.66 & 0.64\\
NorESM & 0.36 & 0.37 & 1234 & 203 & 3.8 & 0.25 & 0.69 & 0.58\\
\hline
\end{tabular}\caption{Comparison of the simulation of rainfall at local and all-India scale by different GCMs for the period 2000-2005}
\end{table}

The results are shown in Table 1. Clearly, we see that most GCMs are not able to represent these characterestics satisfactorily. Most of the models fail to simulate even the mean spatial pattern, as the $spatcr$ is quite low. The standard deviation $SY$ is either overestimated or underestimated by most models, except CNRM-CM5 and Nor-ESM. Only 2 models: CNRM- CM5 and IPSL-CM5 have $X100$ - the number of rain-events above 100mm - close to the true value, while most other models severely underestimate this quantity, though a few overestimate it severely as well. The mean length of wet spells is overestimated by most models, except CMCC-CM5 and MPI-ESM. The temporal correlation of spatial patterns is also highly overestimated by all the models. Most models have very little correlation of daily rainfall with the true values,  with the exception of CESM-CAM5, for whom this correlation is 0.35. Overall, it can be said that GCMs are quite incapable of preserving the spatio-temporal properties of the process.

\section{Model Performance Evaluation}
Now, we come to our main section - the evaluation of our proposed models. For this we consider three settings, as follows.

\subsection{Evaluation of Latent Variables}
In the first setting, we use data for two specific years - 2006 and 2007, and learn the MRF model for these years, over the months June-September. This gives us both the estimate of model parameters like $\alpha,\beta,\mu,\sigma,\tau,\theta,\pi,\lambda$, but also the estimated latent variables $Z^{MRF}$ and $U^{MRF}$. Using the $U^{MRF}$ of these years as the condition for conditional simulation, we now simulate the $Z$-variables for these two years, using the 6 models (Model 1 and Model 2 do not use $U$). This process is repeated for both the low-resolution and the high-resolution datasets. The results are shown in Tables 2-5. 

\begin{table}
\centering
\begin{tabular}{| c | c | c | c | c | c | c |}
\hline
Model & ZZ1 & SCoh & TCoh & SpDiv & nZ1U1 & nZ1U2\\
\hline
MRF-DATA & 13059 & 0.89 & 0.92 & 0.78 & 169 & 60\\
Model1 & 13067 & 0.69 & 0.66 & 0.42 & 107 & 107\\
Model2 & 12727 & 0.7 & 0.93 & 0.58 & 106 & 105\\
Model3 & 13429 & 0.72 & 0.7 & 0.95 & 167 & 66\\
Model4 & 13754 & 0.71 & 0.88 & 0.9 & 164 & 80\\
Model5 & 14795 & 0.77 & 0.88 & 0.91 & 169 & 82\\
Model6 & 14723 & 0.77 & 0.87 & 0.81 & 171 & 82\\
\hline
\end{tabular}\caption{Evaluation of the latent variables simulated by different models with respect to the MRF for the year 2006 on the low-resolution dataset}

\begin{tabular}{| c | c | c | c | c | c | c |}
\hline
Model & ZZ1 & SCoh & TCoh & SpDiv & nZ1U1 & nZ1U2\\
\hline
MRF-DATA & 10729 & 0.87 & 0.89 & 0.86 & 130 & 36\\
Model1 & 10924 & 0.68 & 0.64 & 0.54 & 88 & 86\\
Model2 & 10914 & 0.68 & 0.89 & 0.74 & 93 & 76\\
Model3 & 11422 & 0.68 & 0.63 & 0.91 & 133 & 49\\
Model4 & 11908 & 0.67 & 0.85 & 0.89 & 126 & 62\\
Model5 & 13397 & 0.75 & 0.85 & 0.89 & 136 & 63\\
Model6 & 13234 & 0.75 & 0.85 & 0.79 & 134 & 66\\
\hline
\end{tabular}\caption{Evaluation of the latent variables simulated by different models with respect to the MRF for the year 2007 on the low-resolution dataset}

\begin{tabular}{| c | c | c | c | c | c | c |}
\hline
Model & ZZ1 & SCoh & TCoh & SpDiv & nZ1U1 & nZ1U2\\
\hline
MRF-DATA & 152535 & 0.91 & 0.91 & 0.84 & 1965 & 766\\
Model1 & 152620 & 0.65 & 0.63 & 0.51 & 1247 & 1247\\
Model2 & 150080 & 0.66 & 0.91 & 0.7 & 1243 & 1220\\
Model3 & 159380 & 0.67 & 0.65 & 0.99 & 1898 & 963\\
Model4 & 163910 & 0.66 & 0.87 & 0.97 & 1892 & 1009\\
Model5 & 160480 & 0.85 & 0.88 & 0.98 & 1862 & 967\\
Model6 & 160271 & 0.85 & 0.88 & 0.88 & 1951 & 855\\
\hline
\end{tabular}\caption{Evaluation of the latent variables simulated by different models with respect to the MRF for the year 2006 on the high-resolution dataset}

\begin{tabular}{| c | c | c | c | c | c | c |}
\hline
Model & ZZ1 & SCoh & TCoh & SpDiv & nZ1U1 & nZ1U2\\
\hline
MRF-DATA & 149109 & 0.9 & 0.9 & 0.85 & 1748 & 558\\
Model1 & 149245 & 0.65 & 0.64 & 0.61 & 1221 & 1211\\
Model2 & 147650 & 0.65 & 0.9 & 0.96 & 1228 & 1088\\
Model3 & 156320 & 0.66 & 0.64 & 0.99 & 1761 & 817\\
Model4 & 162350 & 0.65 & 0.66 & 0.99 & 1633 & 896\\
Model5 & 152980 & 0.84 & 0.86 & 0.95 & 1622 & 798\\
Model6 & 152460 & 0.84 & 0.86 & 0.79 & 1667 & 791\\
\hline
\end{tabular}\caption{Evaluation of the latent variables simulated by different models with respect to the MRF for the year 2007 on the high-resolution dataset}
\end{table}

The Tables carry a few broad messages. None of the models are able to achieve the desired level of spatial coherence, but Models 5 and 6 are better than the others (since they utilize the coherent zones). Model 2, which uses location-specific Markov models on $Z$-s is able to achieve perfect temporal coherence, but all other models fall short. The spatial diversity is under-estimated by models 1 and 2, and overestimated by models 3, 4 and 5, while Model 6 is the best in this regard. Models 3 and 6 are closer than the rest to the conditional distribution of local active states, based on the all-India state. However, most models overestimate the number of local active states when the all-India state is 2 ($nZ1U2$). Also, Models 3-6 overestimate the number of local positive states ($ZZ1$).

\subsection{Conditional Simulation with Daily all-India Rainfall}
We next consider conditional simulation, where the model's evaluation period is beyond its training period. The model parameters are learnt using the Markov Random Fields for 6 years- every even year in 2000-2011, for both the high-resoution and low-resolution datasets. Using these parameters, we simulate daily monsoon rainfall for the period 2000-2011 for low-resolution dataset, and 2000-2015 for high-resolution dataset, for the months June-September each year. The simulations are conditioned on all-India rainfall each day, as discussed in Section 7. Clearly, in this case there is no possibility of evaluating the simulated latent variables $Z$ (as they are known only in the training years), so instead we evaluate the simulated $X$ and $Y$ variables, using the criteria discussed in Section 8. The results are shown in Tables 6 and 7.

\begin{table}
\centering
\begin{tabular}{| c | c | c | c | c | c | c | c | c |}
\hline
Model & dMX & dSX & SY & X100 & wetln & dcr & scr & tcr \\
\hline
DATA  &  0  &  0  & 1212 & 2542 & 1.9 & 1 & 0.58 & 0.37 \\
Model1 & 0.08 & 0.1 & 302 & 2595 & 1.3 & 0 & 0.13 & 0.08 \\
Model2 & 0.1 & 0.11 & 309 & 2598 & 1.8 & 0 & 0.13 & 0.28\\
Model3 & 0.23 & 0.17 & 843 & 2648 & 1.4 & 0.8 & 0.16 & 0.1 \\
Model4 & 0.11 & 0.11 & 775 & 2667 & 1.8 & 0.72 & 0.15 & 0.27 \\
Model5 & 0.16 & 0.13 & 849 & 2838 & 1.8 & 0.7 & 0.23 & 0.26 \\
Model6 & 0.1 & 0.17 & 823 & 1776 & 1.9 & 0.63 & 0.4 & 0.3 \\
\hline
\end{tabular}\caption{Evaluation of rainfall simulated by different models, conditioned on daily all-India rainfall for the period 2000-2011, on the low-resolution dataset}
\begin{tabular}{| c | c | c | c | c | c | c | c | c |}
\hline
Model & dMX & dSX & SY & X100 & wetln & dcr & scr & tcr\\
\hline
DATA & 0 & 0 & 1529 & 38831 & 2.7 & 1 & 0.69 & 0.38\\
Model1 & 0.14 & 0.11 & 104 & 37046 & 1.3 & 0 & 0.04 & 0.1\\
Model2 & 0.15 & 0.12 & 131 & 36767 & 1.8 & 0 & 0.04 & 0.3\\
Model3 & 0.14 & 0.11 & 1014 & 38171 & 1.4 & 0.7 & 0.07 & 0.11\\
Model4 & 0.15 & 0.13 & 935 & 38089 & 1.8 & 0.69 & 0.07 & 0.28\\
Model5 & 0.21 & 0.16 & 1030 & 37851 & 1.8 & 0.65 & 0.23 & 0.28\\
Model6 & 0.14 & 0.34 & 979 & 15817 & 1.9 & 0.61 & 0.58 & 0.36\\
\hline
\end{tabular}\caption{Evaluation of rainfall simulated by different models, conditioned on daily all-India rainfall for the period 2000-2015, on the high-resolution dataset}
\end{table}

Once again, the tables reveal some broad patterns of results. First of all, we note that for all the 6 models, the mean and standard deviation parameters of the simulated $X$-variables are much better those of the GCMs presented in Table 1, as their errors compared to the ground truth are much smaller ($dMX,dSX$). Also, all these models are able to perfectly replicate the spatial pattern, due to which the $spatcr$ value is over 0.95 for all models (not shown in tables). However, all models underestimate the standard deviation of $Y$, i.e. the daily total rainfall. Model  6 is grossly underestimates the number of local extreme events which is understandable, as its model construction forces every local $X$ to a fraction of the total rainfall in its zone. The other models are able to simulate this property reasonably. The mean length of wet spells is underestimated by all models except Model 6, which matches the true value on the low-resolution dataset, but falls short on the high-resolution one. The spatial correlation is low for all models compared to the data, though Models 5 and 6 are somewhat better than the other models in this respect. The temporal correlation is best for Model 6. The models 3-6 also show reasonably good daily correlation, which is expected since they are all linked to the actual days by $Y$. However, this correlation is best for Models 3 and 4.

\subsection{Conditional Simulation with local Rainfall}

Next, we come to the second kind of conditional simulation as discussed in Section 7. The $X$-value at a fraction $p$ of all the $S*T$ spatio-temporal  locations are uncovered. Two settings are considered- $p=0.25$ and $p=0.5$. As in the previous case, the model parameters are learnt using MRF using the June-September period of every even year in 2000-2011., while the simulation is done for these months in all the 12 years of this period.  The revealed information is utilized as described in Section 7, followed by simulation by the models. The spatio-temoral locations where the $X$-values are known are excluded from the simulation. The results are shown in Tables 8 and 9. Note that for Model 1 this makes no difference (as all locations are independent for it), while Model 6 is same as Model 5, so these two models are not evaluated.

\begin{table}
\centering
\begin{tabular}{| c | c | c | c | c | c | c | c | c |}
\hline
Model & dMX & dSX & SY & X100 & wetln & dcr & scr & tcr\\
\hline
DATA & 0 & 0 & 1212 & 2542 & 1.9 & 1 & 0.58 & 0.37 \\
Model2 & 0.29 & 0.17 & 702 & 2803 & 1.9 & 0.83 & 0.18 & 0.24\\
Model3 & 0.2 & 0.18 & 463 & 3465 & 1.4 & 0.66 & 0.15 & 0.11\\
Model4 & 0.3 & 0.17 & 726 & 2813 & 1.9 & 0.83 & 0.18 & 0.23\\
Model5 & 0.24 & 0.15 & 956 & 2953 & 1.7 & 0.86 & 0.27 & 0.2\\
\hline
\end{tabular}\caption{Evaluation of rainfall simulated by different models, conditioned on $25\%$ of the spatio-temporal locations for the period 2000-2011, on the low-resolution dataset}

\begin{tabular}{| c | c | c | c | c | c | c | c | c |}
\hline
Model & dMX & dSX & SY & X100 & wetln & dcr & scr & tcr \\
\hline
DATA & 0 & 0 & 1212 & 2542 & 1.9 & 1 & 0.58 & 0.37 \\
Model2 & 0.15 & 0.11 & 921 & 2803 & 1.9 & 0.93 & 0.27 & 0.26\\
Model3 & 0.15 & 0.14 & 677 & 3465 & 1.4 & 0.89 & 0.22 & 0.14\\
Model4 & 0.15 & 0.11 & 944 & 2813 & 1.9 & 0.93 & 0.27 & 0.25\\
Model5 & 0.19 & 0.13 & 1084 & 2953 & 1.7 & 0.94 & 0.34 & 0.21\\
\hline
\end{tabular}\caption{Evaluation of rainfall simulated by different models, conditioned on $50\%$ of the spatio-temporal locations for the period 2000-2011, on the low-resolution dataset}
\end{table}

This setting clearly benefits Model 2, as it is now able to get some input in form of the $Z$-variables. Clearly this improves its performance with respect to daily correlation and spatial correlation. Daily correlation for all models are found to increase. The tables clearly show that increasing the number of observations from $p=0.25$ to $p=0.5$ improves the performances of all models. Local information with $p=0.25$ gives a somewhat poorer performance with respect to local statistics ($dMX,dSX$) compared to daily all-India rainfall, but with $p=0.5$ these are improved.

\subsection{Unconditional Simulation}
Finally, we carry out our last experiment- where the models are run without any contextual information. As before we learn the model parameters from every even year in 2000-2011, and run them for 12 years without any input. The results for the low-resolution and high-resolution data are shown in Tables 10 and 11. Since daily correlation is irrelevant in this case (as there is no connection with the actual days) we drop that criteria from these tables. It can be seen that the figures are hardly different from those in case of conditional simulations. in Tables 6-9. This shows that our models can achieve decent simulation performance without any external supervision.

\begin{table}
\centering
\begin{tabular}{| c | c | c | c | c | c | c | c |}
\hline
Model & dMX & dSX & SY & X100 & wetln & scr & tcr\\
\hline
DATA   &  0  &  0  & 1212 & 2542 & 1.9 & 0.58 & 0.37 \\
Model1 & 0.08 & 0.1 & 302 & 2595 & 1.3 & 0.13 & 0.08 \\
Model2 & 0.1 & 0.11 & 309 & 2598 & 1.8 & 0.13 & 0.28\\
Model3 & 0.24 & 0.16 & 846 & 2523 & 1.4 & 0.16 & 0.09\\
Model4 & 0.11 & 0.11 & 729 & 2472 & 1.8 & 0.15 & 0.28\\
Model5 & 0.16 & 0.13 & 819 & 2690 & 1.8 & 0.23 & 0.27\\
Model6 & 0.1  & 0.17 & 839 & 1961 & 1.8 & 0.41 & 0.3\\
\hline
\end{tabular}\caption{Evaluation of rainfall simulated unconditionally by different models, for the period 2000-2011, on the low-resolution dataset}
\begin{tabular}{| c | c | c | c | c | c | c | c |}
\hline
Model & dMX & dSX & SY & X100 & wetln & scr & tcr\\
\hline
DATA & 0 & 0 & 1529 & 38831 & 2.7 & 0.69 & 0.38\\
Model1 & 0.14 & 0.11 & 104 & 37046 & 1.3 & 0.04 & 0.1\\
Model2 & 0.15 & 0.12 & 131 & 36767 & 1.8 & 0.04 & 0.3\\
Model3 & 0.15 & 0.12 & 1027 & 33742 & 1.4 & 0.08 & 0.12\\
Model4 & 0.17 & 0.13 & 936 & 33272 & 1.7 & 0.08 & 0.29\\
Model5 & 0.18 & 0.14 & 1036 & 32158 & 1.7 & 0.24 & 0.28\\
Model6 & 0.13 & 0.27 & 972 & 12885	 & 1.8 & 0.58 & 0.37\\
\hline
\end{tabular}\caption{Evaluation of rainfall simulated unconditionally by different models, for the period 2000-2015, on the high-resolution dataset}
\end{table}

\subsection{Conclusions from the experiments}
From the experiments discussed above, a few points are worth noting. First of all, we find that the GCMs perform poorly with respect to most of the criteria discussed here. Among the different models we considered here, there is no single model which can be considered as outstanding. Model 6 uses the least number of parameters and performs best with respect to spatial and temporal correlations, even though it falls short of the ground-truth. It also simulates the spatial diversity (SpDiv) better than other models. However, it greatly underestimates the number of extreme rainfall events. It is also unable to capture the local statistics very well, as $dSX$ is high for it in all the settings. Also, models 5 and 6 tend to overestimate the number of locations in active state. Among other models, Model 2 works reasonably well for conditional simulation based on local information, and it is able to simulate the temporal coherence reasonably well. It also makes reasonable estimate of the local and all-India statistics (dMX,dSX,SY). Model 3, on the other hand, works better for conditional simulation based on daily all-India rainfall, though it is unable to simulate the spatial and temporal correlations. Model 4 is aimed to be compromise between Model 2 and Model 3, and its simulation results show this. It can simulate the temporal coherence well, and it has high daily correlations for both kinds of conditional simulation. However, this is compenstated by its high parameter complexity, and it cannot simulate spatial correlation either. Note that Models 1-5 make a reasonable estimate of the number of local extreme rainfall events, though none of them is specially equipped to do so. Model 5 is intermediate in all respects as it simulates all the criteria to some extent, though not particularly well.

In short, the Indian monsoon rainfall is a very complex phenomena, so that it is very difficult to simulate all of its properties simultaneously. Simulating one set of properties well results in poor simulation of some other properties.

\section{Possible Extensions and Conclusions}
In this work, we attempted to build stochastic rainfall generators for India, that can preserve the complex spatio-temporal characteristics of the phenomenon. Unlike other contemporary approaches to stochastic rainfall simulation, we made use of a Markov Random Field to estimate location-specific rainfall parameters to maintain spatial smoothness, and we also took a Bayesian non-parametric approach to demarcate the entire landmass into spatially coherent zones based on daily local conditions. Using these, we were able to partially achieve the spatial correlation of rainfall in our simulations. We proposed a large number of criteria to evaluate the models with respect to local statistics and spatio-temporal properties, and compared the merits and demerits of different models. We also showed that General Circulation Models are not at all satisfactory with respect to these properties. We concluded that preserving all these properties simultaneously in a simulation is quite challenging. Finally, we also showed how our models can incorporate some external information to improve their simulations and perform conditional simulation.

There are a number of directions along which this work can be extended. In most stochastic weather generators, rainfall simulation is the most important step, but this is in turn used to simulate more climatic variables such as temperature, and we can also do so. Secondly, these simulators also attempt to simulate rainfall at very high resolutions- where no observed data is available. We will extend our models to achieve this, based on our capability of conditional simulation.  Also, we aim to consider more sophisticated hierarchical approaches to improve spatial correlations and simulation of extreme events.